# Visitors to urban greenspace have higher sentiment and lower negativity on Twitter

Aaron J. SCHWARTZ, [a,b,c]; Peter SHERIDAN DODDS, [b,d]; Jarlath P.M. O'NEIL-DUNNE [a,c]; Christopher M. DANFORTH [a,b,d]; Taylor H. RICKETTS [a,c];

**Affiliations:**
a. Gund Institute for Environment, University of Vermont, Burlington, VT 05405, USA
b. Vermont Complex Systems Center, University of Vermont, Burlington, VT 05405, USA
c. Rubenstein School of Environment and Natural Resources, University of Vermont, Burlington, VT 05405, USA
d. Department of Mathematics and Statistics, University of Vermont, Burlington, VT 05405, USA

## Abstract

1. With more people living in cities, we are witnessing a decline in exposure to nature. A growing body of research has demonstrated an association between nature contact and improved mood.
2. Here, we used Twitter and the Hedonometer, a world analysis tool, to investigate how sentiment, or the estimated happiness of the words people write, varied before, during, and after visits to San Francisco's urban park system. We found that sentiment was substantially higher during park visits and remained elevated for several hours following the visit.
3. Leveraging differences in vegetative cover across park types, we explored how different types of outdoor public spaces may contribute to subjective well-being. Tweets during visits to Regional Parks, which are greener and have greater vegetative cover, exhibited larger increases in sentiment than tweets during visits to Civic Plazas and Squares.
4. Finally, we analyzed word frequencies to explore several mechanisms theorized to link nature exposure with mental and cognitive benefits. Negation words such as 'no', 'not', and 'don't' decreased in frequency during visits to urban parks.
5. These results can be used by urban planners and public health officials to better target nature contact recommendations for growing urban populations.

# 1. Introduction

There is a growing interest in understanding the connection between mental health and exposure to biodiversity, due to the simultaneous growth of urban areas globally and rising rates of mood disorders (Murray et al., 2012; United Nations, 2014). While cities provide access to significant economic and social opportunities, researchers have identified an urban health penalty that arises from the pace of life, exposure to environmental stressors and chemicals, and disconnect from diverse natural environments in which human evolution occurred (Bettencourt, Lobo, Helbing, Kuhnert, & West, 2007; McDonald, Beatley, & Elmqvist, 2018). Urban greenspace, and specifically urban parks, are a policy instrument that can help reduce the impacts of "nature deficit disorder" (Louv, 2011).

Studies on the mental benefits of nature exposure have typically taken one of two approaches. First, broad studies based on surveys and geographic data have established an association between proximate natural areas and subjective well-being (Hartig, Mitchell, de Vries, & Frumkin, 2014; Maas, Verheij, Groenewegen, de Vries, & Spreeuwenberg, 2006; van den Berg et al., 2016; Wheeler et al., 2015). The Normalized Difference Vegetation Index (NDVI), a proxy for vegetation derived from remotely sensed data, has been used as a measure of neighborhood greenness and is associated with lower levels of depression (Fong, Hart, & James, 2018). High levels of cumulative childhood greenspace exposure were associated with lower risk of developing psychiatric disorders (Engemann et al., 2019). Broad surveys are unable to capture acute exposure events to greenspace and biodiversity, making it challenging to identify the types of natural areas most effective at delivering mental benefits (Bell, Phoenix, Lovell, & Wheeler, 2014). Field experiments address this issue by directly exposing participants to natural areas. For example, participants walking through natural areas showed improved affect and cognition compared to those walking through urban environments (Bratman, Daily, Levy, & Gross, 2015). In another experiment, individuals who walked in areas with greater biodiversity reported higher levels of subjective well-being (Carrus et al., 2015). Several recent experiments have examined landscape preferences, landscape structure, and biodiversity across a gradient of park types with larger experimental pools (Fischer et al., 2018; Hoyle, Hitchmough, & Jorgensen, 2017; Hoyle, Jorgensen, & Hitchmough, 2019; Martens, Gutscher, & Bauer, 2011; Qiu, Lindberg, & Nielsen, 2013). Here, we combined the strengths of these approaches by following a large group of people making known visits to a range of park types.

Recently, mobile phone applications have been used to conduct *ecological momentary assessments,* querying users about their mood and environment in real-time (Bakolis et al., 2018; MacKerron & Mourato, 2013; McEwan et al., 2019). In the present study, we use location-enabled data from social media to observe individuals at a level of geographical precision that indicates actual contact with greenspace and biodiversity. Previous studies analyzing tweets in urban greenspace have studied comparative well-being across a city, emotional changes across seasons, and different ways of analyzing the emotional content of tweets (Lim et al., 2018; Plunz et al., 2019; Roberts et al., 2018).

We used the Hedonometer, a word analysis tool that quantifies the sentiment of text based on the happiness values attributed to English words (Dodds & Danforth, 2010; Dodds et al., 2011). The Hedonometer has been demonstrated to correlate with traditional survey metrics of subjective well-being at the city and state level, including Gallup's well-being index and United Health Foundation's America's health ranking (Mitchell, Frank, Harris, Dodds, & Danforth, 2013). The Hedonometer has also been deployed to analyze the discourse around climate change following hurricanes (Cody et al. 2015). Using the Hedonometer's sentiment dictionary we asked: *(Q1) What is the magnitude and duration of the change in sentiment from visiting urban parks?*

Using geo-located tweets allowed us to differentiate between different doses of nature exposure intensity, defined as the quality and quantity of nature itself, as called for in prior work (Shanahan et al., 2016). The San Francisco Recreation and Parks Department classifies their facilities into categories based on park size, design, and amenities. Using official park types along with estimates of park vegetative cover, we investigated how different types of nature exposure are associated with changes in happiness as expressed in tweets. While we don't measure biodiversity directly, we use NDVI and vegetative cover as proxies for the intensity of nature exposure. We asked: *(Q2) What is the association of park type and vegetative cover with the change in sentiment from park visitation?*

Complementary theories from psychology and neurobiology suggest several mechanisms connecting nature exposure with mental state (Berto, 2014; Frumkin et al., 2017). The Biophilia hypothesis suggests that humans have an innate affinity for natural environments similar to those in which we evolved (Kellert & Wilson, 1995). More specifically, Stress Reduction Theory (SRT) predicts a decrease in physiological stress following nature contact, resulting in a variety of positive health outcomes (Ulrich et al., 1991; Ward Thompson, Aspinall, Roe, Robertson, & Miller, 2016). Attention Restoration Theory (ART) predicts that time in nature provides the opportunity to restore directed attention capacity, which results in improved cognition (S. Kaplan, 1995; Ohly et al., 2016). Nature exposure has also been found to correlate with increased prosocial behavior through 'unselfing', a shift away from self-interest and towards generosity (Zhang, Piff, Iyer, Koleva, & Keltner, 2014). A recent review of the pathways linking greenspace to health called for quasi-experimental studies and the assessment of varying exposure types to better explore the mechanisms underlying the mental benefits of nature contact (Markevych et al., 2017). Here, we analyzed word frequency patterns around park visitation to explore the mechanisms driving mental shifts from park visitation. We derived word frequency time series from tweets and asked: *(Q3) What do word frequency patterns indicate about the mechanisms driving the change in sentiment from park visitation?*

**2. Methods**
*2.1. Study Site & Data Collection*
Using Twitter's streaming Application Programming Interface ("Twitter Streaming API," 2019), we collected all tweets explicitly geotagged with latitude and longitude originating in the San Francisco, USA (2016 Population Estimate: 871,000) area between May 19, 2016 and August 2, 2016 (roughly 70,000 tweets per day). Although Twitter places a rate limit on the streaming API, our sample came from a relatively small geographic area leading to insignificant errors for tweet collection. We selected San Francisco as a study site due to its diverse park system, which spans more than 220 sites and 3,400 acres. According to the Trust for Public Land, 98.2% of San Francisco's population live within walking distance of a park and San Francisco has one of the top ranked park systems in the USA (Harnik, Martin, & Barnhart, 2015). Using the *Python* (v.2.7) geographic libraries *Fiona* (v. 1.5.1) and *Shapely* (v. 1.6), we determined which tweets fell within San Francisco Recreation and Parks Department facility boundaries ("Park and Open Space Map," 2016). Finding tweets inside parks depends on the accuracy of mobile GPS; some of our user pool may have been just outside of parks due to measurement error. San Francisco Recreation and Parks categorizes their facilities into nine categories, with 94% of Tweets collected located in the following three categories: Regional Parks, Civic Plazas and Squares, and Neighborhood Parks and Playgrounds (Fig. 1). These parks were categorized by a professional parks planner according to guidelines determined by San Francisco Recreation and Parks Department.

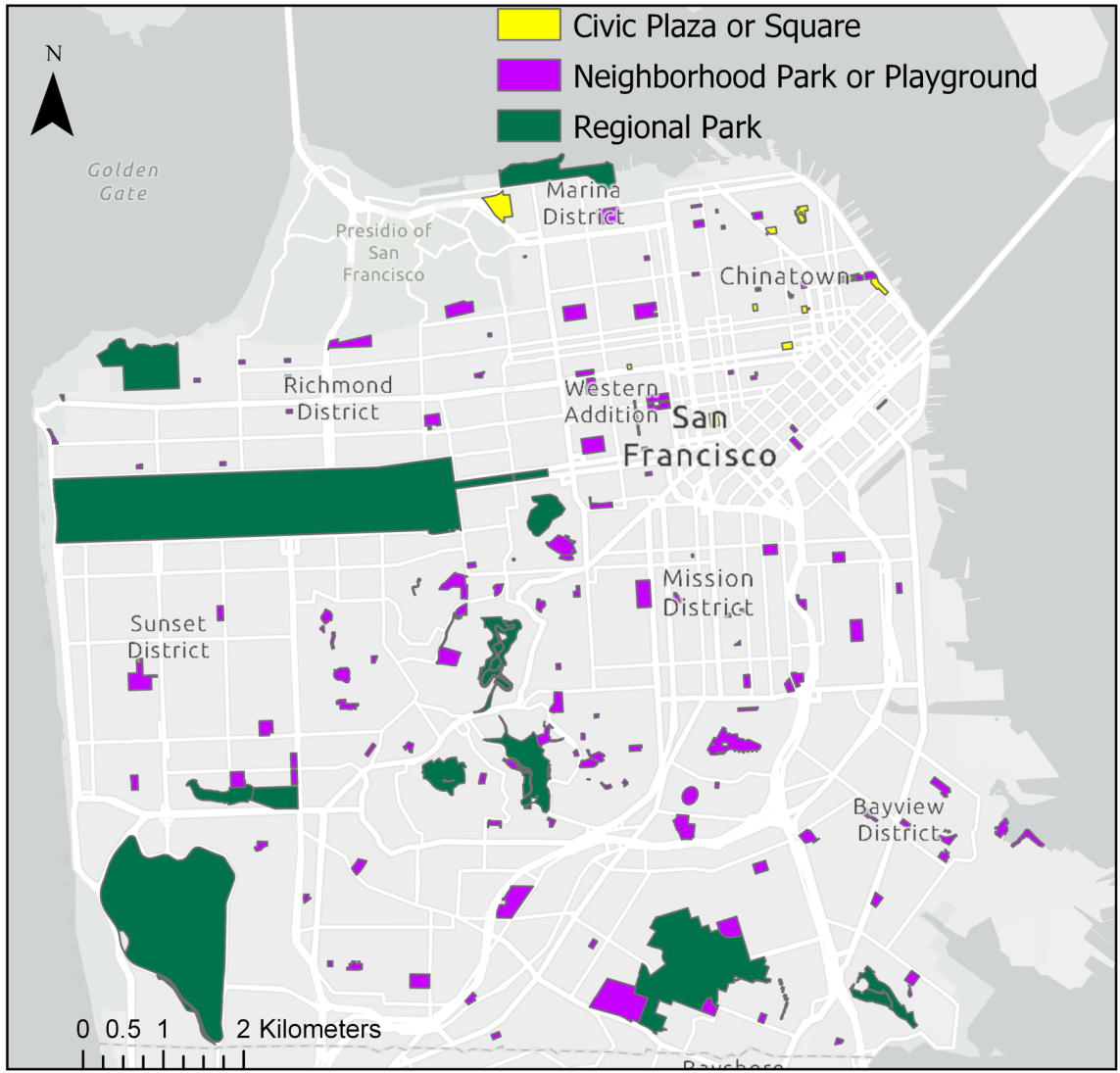

*Figure 1. San Francisco Recreation and Parks Facility Map for 3 park categories.*

We constructed a list of Twitter users who had visited at least one park during the study period and used the Twitter API to download their 3,200 most recent tweets. A month later, we updated user histories with any tweets posted since the park visit. We used several heuristics to remove automated bots and businesses from the user sample and additionally removed any individual who made their account private in the period following their park tweet. We also removed users who did not have a park visit tweet in English. Our process resulted in 4,688 user timelines.

*2.2. Tweet Binning*

We saved the following fields for each tweet within a user's timeline: message identification string, timestamp, text, language, and location. We used tweet timelines as the raw data for all further analysis. We defined a park exposure as the first tweet posted from within a park on a given calendar day. We assigned all other tweets as "pre" or "post" to the closest park exposure tweet before or after, enabling the binning of tweets across users into hourly bins. For example, if a user tweeted in a park a 2PM, and also tweeted at 10:30 AM and 4:15 PM, the user would have tweets in the -4, 0 (in-park), and +3 bins. If we encountered subsequent tweets that also occurred in parks on a given day, we treated them the same as all other post-park tweets. This avoids the bias of including several consecutive park tweets in the park exposure bin and simplifies the assignment of out of park tweets to the initial exposure. Users had an average of 0.62 in-park tweets within 24 hours of their initial park exposure tweets and 78% of users had no secondary park tweets. By pooling users into relative time bins, we were able to create large enough word samples to apply sentiment analysis.

*2.3. Sentiment Analysis*

The Hedonometer includes a sentiment dictionary for 10,022 of the most commonly used English words, merged from four distinct text corpora. The Hedonometer performs favorably compared with other sentiment dictionaries, using a continuum scoring of words with high coverage (Reagan, Danforth, Tivnan, Williams, & Dodds, 2017). Word ratings were calculated by averaging scores from a pool of online crowdsourced workers at Amazon's Mechanical Turk (Dodds et al., 2011). The words were rated on a scale from 1 (least happy) to 9 (most happy). For example, 'sunshine' has a score of 7.9 and 'traffic' has a score of 3.3. Words with scores between 4 and 6 were excluded from the analysis either because they are emotionally neutral (e.g., 'at' [4.9], 'and' [5.2]) or because they are context dependent (e.g. 'church' [5.5], 'capitalism' [5.2]). For our study, we also removed any words appearing in the names of San Francisco Parks from the analysis (e.g., 'golden' [7.3], 'gate' [5.1], and 'park' [7.1]). We recognize that words representing natural features such as 'beach' [7.9], 'tree' [7.1], 'grass' [6.5] typically have positive sentiment and hypothesize that the presence of such words indicates awareness of the surrounding environment, which has been connected with a reduction in stress (Ulrich et al., 1991). While the Hedonometer does not take word context or order into account, prior use of the tool has indicated that with a sufficiently large sample size (>1000 words), a reliable estimate of text happiness is possible (Reagan et al., 2017). For this reason, we did not measure the happiness of individual tweets or users but instead implemented the pooling procedure described below.

*2.4. Estimating Sentiment*

For a set of tweets, we estimated **sentiment** as the weighted average of word scores using their relative frequencies as weights (Equation 1). We generated sentiment time curves by applying the Hedonometer to hourly bins of tweets before, during, and after park exposure. To provide additional statistical support to this approach, we used a bootstrapping procedure. For a given hourly bin, we randomly selected 80% of the tweets without replacement and calculated the pooled sentiment. Performing this procedure 100 times, we derived a range of plausible mean sentiment values for each tweet bin.

$$\text{Equation 1: Sentiment} = \frac{\sum_{i=1}^{n} v_i f_i}{\sum_{i=1}^{n} f_i}$$

Where $v_i$ is the sentiment score of the $n^{th}$ word and $f_i$ is its frequency in a given text (set of tweets)

To quantify the **change in sentiment** from exposure to urban greenspace, we compared the sentiment from tweets before park visits and during park exposure. First, we defined a set of baseline tweets. For a given park, these were tweets occurring more than 1 and up to 6 hours prior to tweets posted from that park. We subtracted the baseline sentiment from that of the park exposure tweets. To estimate a plausible interval for change in sentiment, we performed a similar bootstrapping procedure. We selected a random 80% of tweets from both the baseline and park tweets and calculated the difference in their sentiment scores. Performing this operation 100 times, we were able to estimate a mean, variance, and a 95% confidence interval for the mean change in sentiment. Robustness checks were performed to show convergence of this range at 100 runs.

*2.5. Duration Calculation*

To estimate the **duration** of a change in sentiment from visiting a park or set of parks, we defined the baseline set of tweets in the same manner as above. We then performed the following bootstrapping procedure in an iterative manner. We started with the tweets occurring one hour after park exposure and estimated the change in sentiment between the baseline and that hourly bin of tweets. We continued to the next bin if we were able to reject the null hypothesis from the one sample T-Test that the mean of the bootstrapped differences is equal to 0 at the 95% confidence level. The duration of a change in sentiment was the last hourly bin at which we are able to reject this null hypothesis. We performed this analysis with and without in-park tweets that occur after initial park tweets on a given day.

*2.6. Park Classifications and NDVI*

To understand how park type relates to the benefits of park exposure, we used the San Francisco Recreation and Parks Department classifications for the 160 parks in which we found tweets during the study period. The vast majority of park acreage and tweet activity occurred in 3 categories: Civic Plaza or Square, Neighborhood Park or Playground, and Regional Park (Fig.

1). We grouped tweets posted from each of these park categories to compare the changes in sentiment from baseline across categories.

We also calculated Normalized Difference Vegetation Index (NDVI) for each of the 160 parks in which tweets occurred. We developed an automated method to map vegetation throughout San Francisco using an object-based approach with 1-meter, 4-band National Agricultural Imagery Program (NAIP) data acquired in the summer of 2016 (O'Neil-Dunne, Pelletier, MacFaden, Troy, & Grove, 2009). We segmented NAIP imagery into image objects using a multiresolution segmentation algorithm (Benz, Hofmann, Willhauck, Lingenfelder, & Heynen, 2004). We computed NDVI for each image object based on the mean near-infrared and red values in the NAIP data (Equation 2).

Equation 2: $NDVI = (NIR - RED) / (NIR + RED)$
(NIR = near-infrared; RED = visible red).

Using a series of classification and morphology algorithms, we assigned objects to one of two classes: vegetation or non-vegetation. We overlaid these objects, along with their NDVI values, onto the San Francisco park polygons to calculate the percent area with vegetation and mean NDVI for each park, excluding pixels defined as bodies of water from data provided by the SF Department of Public Works (*data.sfgov.org*, 2019). NDVI scores range from -1 to 1 with higher scores being greener. We report NDVI and Percent Vegetation for the 3 main park categories in Table 1. The other park categories (Community Gardens, Concessions, Family Camp, Mini parks, Parkways, and Zoological Gardens) were not large enough to accurately estimate mean NDVI or sentiment based on the number of tweets posted from those spaces.

*Table 1. Primary Park Category Characteristics*

| Category | Count | Mean Acres | Mean NDVI | Mean Percent Vegetated |
|---|---|---|---|---|
| **Regional Park** | 13 | 609.37 | 0.21 | 79.48% |
| **Neighborhood Park or Playground** | 112 | 11.54 | 0.12 | 63.44% |
| **Civic Plaza or Square** | 10 | 8.79 | 0.06 | 45.42% |

## 3. Results

*3.1. What is the magnitude and duration of the change in sentiment from visiting urban parks?*
Tweets posted within parks have a higher sentiment than tweets posted before or after park visits. We depict the sentiment time curve for all users in Fig. 2, with average sentiment fluctuating between roughly 6.1 and 6.2 outside of park visits. Sentiment reaches 6.43 across all tweets occurring in parks. The immediate hours before and after park exposure also elevated from baseline. The bootstrapped intervals for mean sentiment are narrower around the park exposure because our dataset contains more tweets during those hours than in any individual hour preceding or following the park exposure.

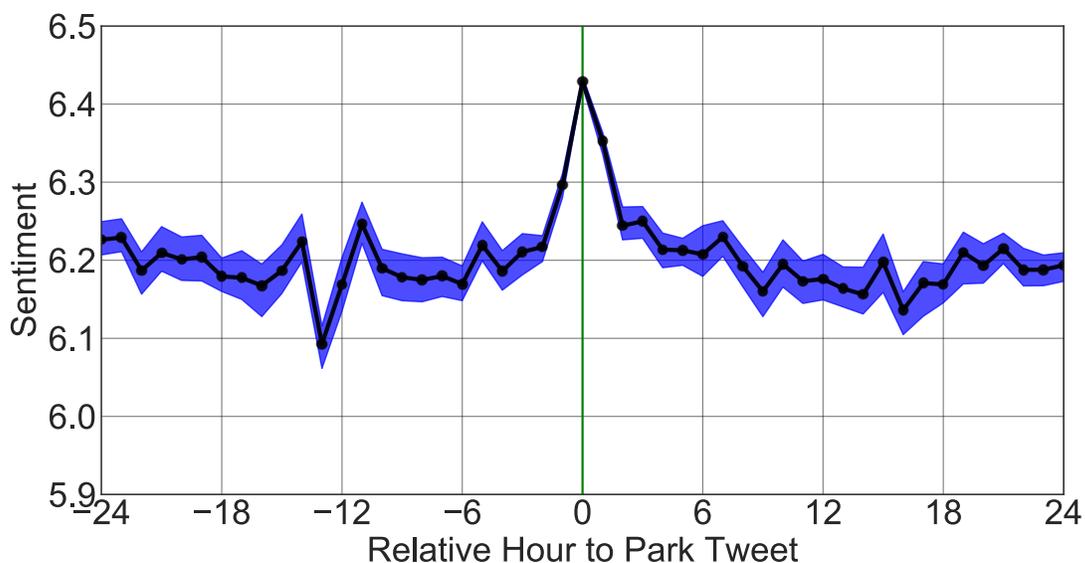

*Figure 2. Sentiment before, during, and after park visit. Average sentiment for all user tweets (y-axis), within 24 hours of park exposure, binned by relative hour to in-park tweet (x-axis). The green vertical line represents the tweet in a San Francisco Park, with highest sentiment value. The blue range is full sentiment range from 100 runs of randomly sampling 80% of tweets.*

The mean change in sentiment for all parks is 0.229 (Bootstrapped 95% CI 0.220, 0.238) (Fig. 3). As a point of reference, the average day on Twitter in 2016 had a sentiment of 6.04, and Christmas Day was the happiest day in 2016 with a sentiment of 6.26 (Hedonometer.org, 2019). Thus, across our user pool, tweets during visits to urban parks exhibited a similar increase in sentiment as the jump on Christmas Day for Twitter as a whole.

Across all parks, we estimate the duration of elevated sentiment after initial park tweets. We find that sentiment remains elevated for four hours, compared to a baseline level averaged over the six hours before park visitation. This analysis included tweets inside parks that occurred after the initial exposure to avoid bias of highly active users and to clarify assignment of post park tweets to an initial park exposure (see methods). To calculate an even more conservative estimate of elevated sentiment, we repeated this analysis without those tweets, resulting in an estimated duration of one hour. We thus expect the duration to fall somewhere between one and four hours.

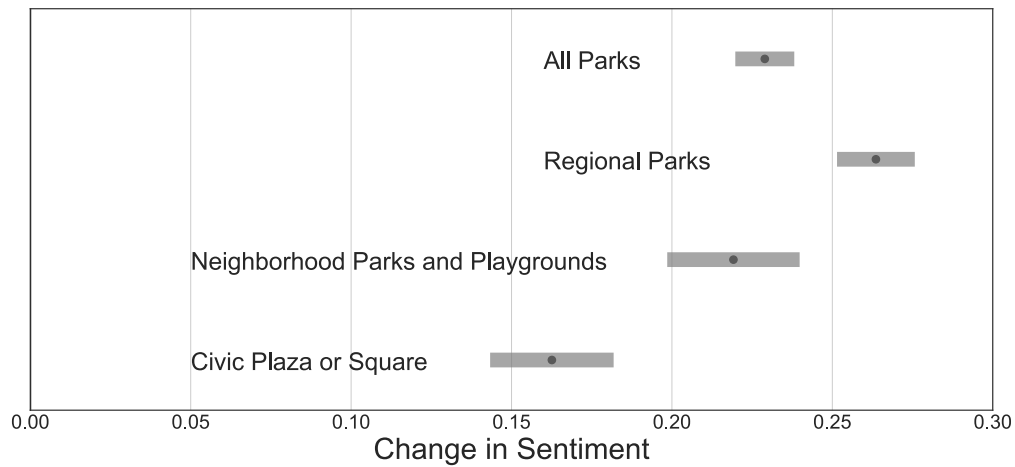

*Figure 3. Comparing change in sentiment between park categories. The horizontal axis is estimated change in sentiment between baseline and in-park tweets. Ranges are 95% Confidence Intervals from 100 runs of bootstrap process. Dots are mean change in sentiment. Park categories are as defined by San Francisco Recreation and Parks Department.*

3.2. What is the association of park type and vegetative cover with the change in sentiment from park visitation?
Regional Parks exhibit the highest mean change in sentiment of 0.264 (0.251, 0.276). Neighborhood Parks and Playgrounds have a moderate mean change in sentiment of 0.219 (0.199, 0.240). Civic Plaza or Squares have the lowest change in sentiment of 0.163 (0.143, 0.181) (Fig. 3). Confidence interval limits do not overlap for any pair of park types, indicating significant differences among them. The differences in mean sentiment among park types correspond to differences in size, NDVI and vegetative cover (Table 1).

3.3. W*hat do word frequency patterns indicate about the mechanisms driving the change in sentiment from park visitation?*
Tweets in parks have higher sentiment than tweets prior to park visitation due to positive words with higher frequency, such as 'beach', 'beautiful', 'festival', 'happy', 'young', 'fun', and negative words with lower frequency, such as 'not', 'no', 'don't', 'can't', and 'wait' (Fig. 4). Of specific interest, negation words such as 'not' and 'don't' fluctuate before and after park exposure but exhibit a marked drop (45% and 47%) in and around the park exposure (Figs 5A-5B). The word 'beautiful' exhibits the opposite pattern, fluctuating around a baseline and then roughly doubling in frequency during park exposure (Fig. 5C). Finally, we examine the first-person word "me" which has a neutral sentiment (and is not included in the sentiment scores above). Use of "me" drops 38% from its mean use level during park visits (Fig. 5D).

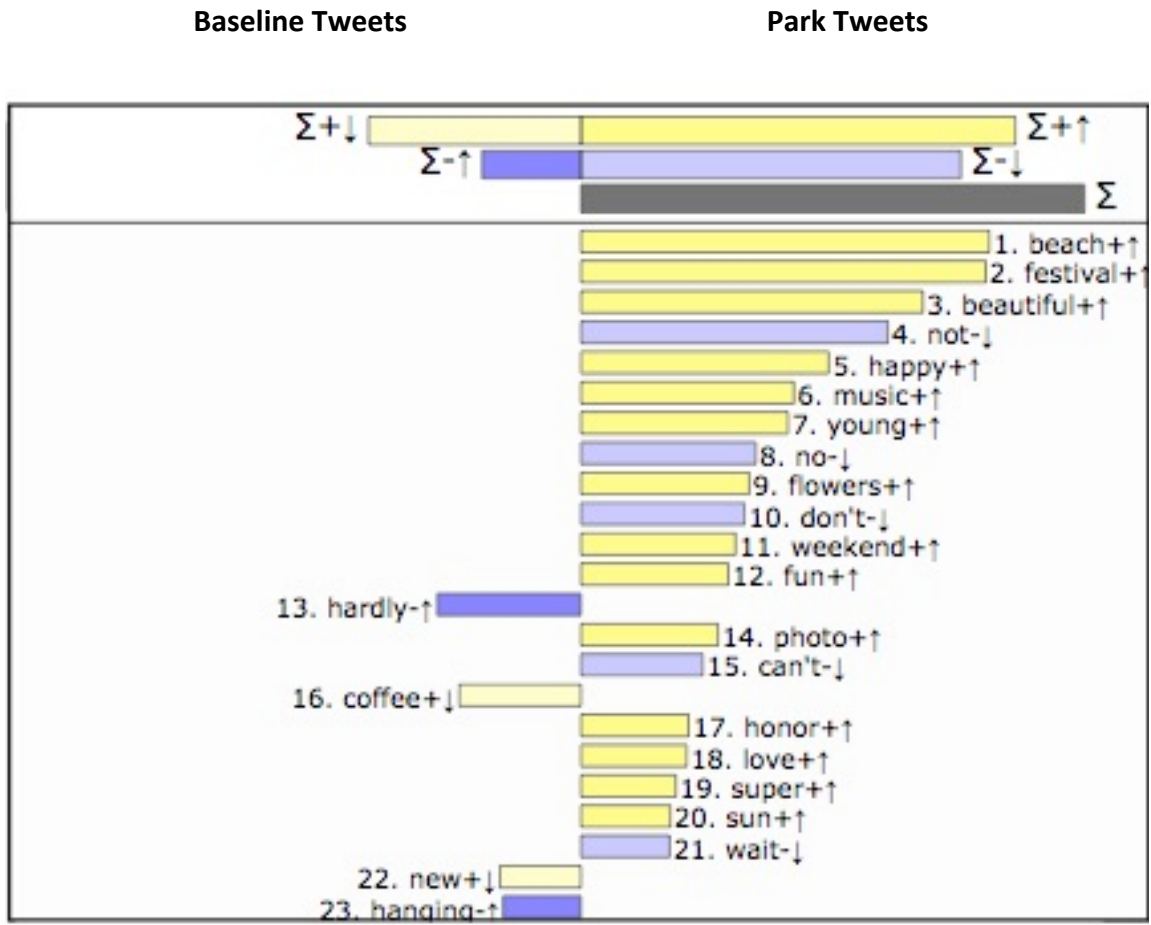

*Figure 4. This figure shows the words driving the difference between park and baseline tweets, in order of decreasing contribution to the difference in sentiment. The right side represents the park tweets, with a mean sentiment of 6.43. The left side represents tweets 1-6 hours preceding the park tweets, with a mean sentiment of 6.20. Purple bars represent words <= 4 (with – symbol) on the Hedonometer scale. Yellow bars represent words >= 6 (with + symbol) on the Hedonometer scale. Arrows indicate whether a word was more or less frequent within that set of tweets compared to the other text. For example, "Beach" is a positive word (purple) with higher frequency in park tweets that contributes to increased sentiment of that set. "Not" is a*

*negative word (purple) that appears less frequently in that set, resulting in a higher relative sentiment score compared to baseline.*

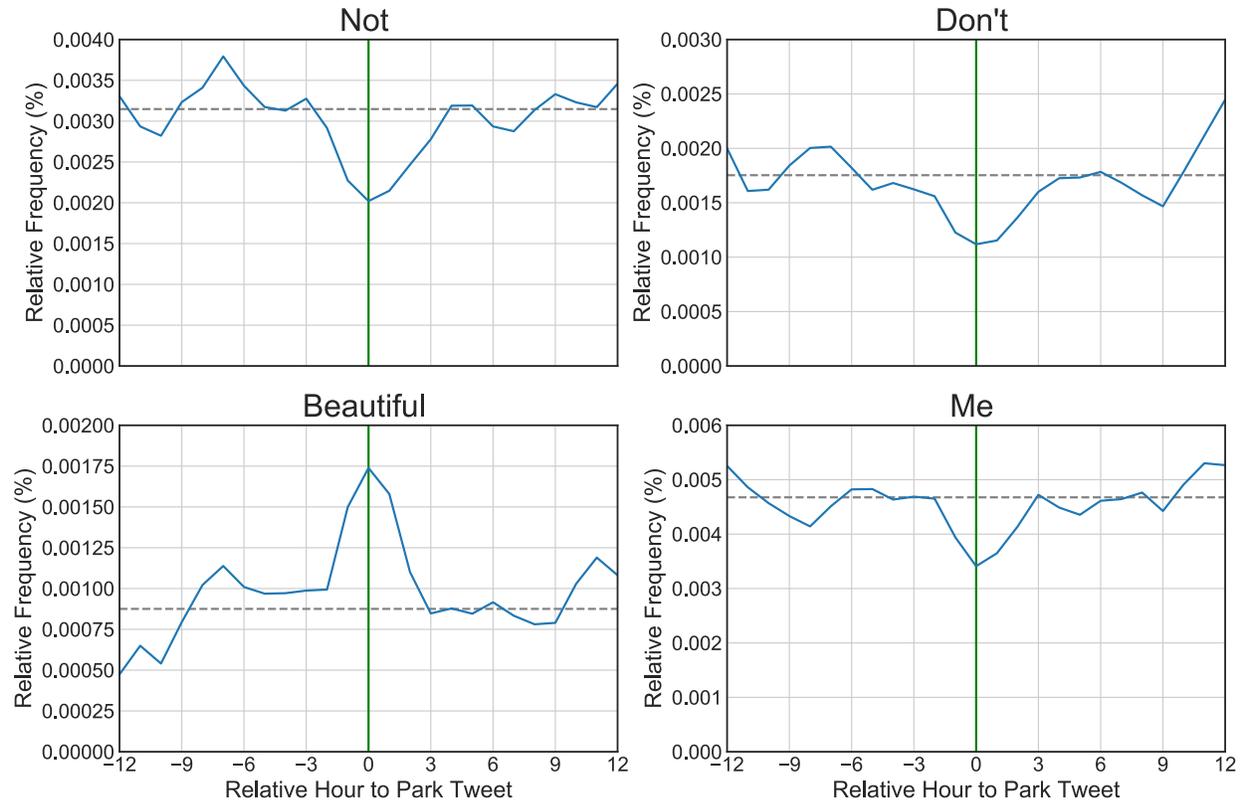

*Fig. 5A-D. Word frequency patterns before and after park visit. X-axis depicts hourly tweet bins from 12 hours before to 12 hours after in-park tweet, which is represented by green line. Y-axis ranges are scaled for each word's relative frequency. Relative frequencies (blue-lines) are smoothed as moving averages over 3 hours. Grey dashed line is mean frequency for entire 24-hour period around park visit*

**4. Discussion & Conclusion**

In our study, tweets posted from urban nature were happier by roughly 0.23 points on the Hedonometer scale from baseline. This increase in sentiment is equivalent to that of Christmas Day for Twitter as a whole in the same year (Hedonometer.org, 2019). Our analysis of duration suggests that elevated sentiment lasts for between 1 and 4 hours following an initial park tweet. The recent *Urban Mind* study found a similar duration for their weeklong study on roughly 100 users self-reporting their happiness in different environments (Bakolis et al., 2018). Interestingly, sentiment begins to increase from baseline in the hour preceding the in-park tweets (Fig. 2). Possible explanations for this trend are anticipation for the park visit, meeting friends on the way, or perhaps relief due to leaving work and heading to a more enjoyable location. Recent work found that the emotion of anticipation increases in greenspaces (Lim et al., 2018); further investigation is merited to better understand the temporal dynamics of anticipation and its relationship with nature contact.

Tweets located in Regional Parks exhibited the strongest increase in sentiment followed by tweets in Neighborhood Parks and Playgrounds and then Civic Plazas and Squares. There are several plausible explanations for the greater sentiment increase occurring in Regional and Neighborhood Parks. Regional Parks have greater vegetative cover and may offer more opportunities for nature contact and exposure to biodiversity compared to Civic Plazas and Squares (Table 1). The greater vegetative and floral diversity found in the larger Regional Parks may be playing a role as indicated by 'flowers' appearing in Figure 4, supported by prior research on the most salient features of landscapes (Hoyle et al., 2017). Alternatively, the large size of Regional Parks may be providing greater restorative capacities through a more distinct separation from the urban environment. Neighborhood Parks and Civic Plazas are close in size but also exhibit a significant difference in sentiment increase, suggesting that park size is not the only factor at play. The three park classes offer different amenities and activity types, which may also be contributing to the differences in sentiment. A recent review summarized the range of scales at which biodiversity can be measured inside of parks – from vegetated versus non-vegetated to genetic diversity – and suggested several directions for future research on the people-biodiversity interface (Botzat, Fischer, & Kowarik, 2016).

The roles of exercise and socialization can be difficult to separate from the direct contributions of nature to enhanced subjective well-being (Ambrey, 2016). Regional Parks may be more amenable to physical activity, although our analysis of words does not suggest that physical activity related terms are driving the elevated happiness of the in-park tweets (Fig. 4). Technologies such as heart-rate monitors may provide new opportunities for distinguishing the benefits of outdoor exercise from the benefits of nature exposure per se. Differences in social interactions across park types may also be contributing to variation in sentiment. Civic Plazas, which tend to be paved and more centrally located, represent an outdoor, public gathering space where people go to socialize in their time away from work. Our results indicate that Regional and Neighborhood Parks are more restorative spaces than Civic Plazas, and that nature *per se* is potentially playing a role in delivering mental benefits to park visitors. While we are unable to measure how much time is spent in a park following a tweet, it is plausible that visits to Regional Parks are longer than visits to the other park classes. Alternative approaches such as Ecological Momentary Assessments or time use surveys may be more effective at capturing the duration of park visits. Recent work has suggested that at least 120 minutes of weekly nature exposure lead to enhanced self-reported health and well-being (White et al., 2019). Future analyses should also look to directly compare nature contact with indoor activities (e.g., museum visits), but these comparisons are beyond the scope of this paper. Several reviews have summarized the growing body of work on nature contact and set a research agenda for building a more nuanced understanding of the relationship between nature contact and health (Frumkin et al., 2017; Hartig et al., 2014).

The mechanisms through which urban nature exposure improves mental health are still being investigated. Green Mind Theory, a recent synthesis of proposed pathways, suggests that the negativity bias of the brain – which may have been evolutionarily advantageous – is constantly activated by the stressors of modern life (Pretty, Rogerson, & Barton, 2017). In our analysis, park visitation coincides with a decrease in words such as 'no', 'don't', and 'never' (Fig. 4). These words, known as negations, are associated with focused, analytical thinking (Pennebaker, 2011). The decrease in negation frequency may provide support for Attention Restoration Theory, which links nature exposure with the experience of soft fascination and can result in improved cognition (R. Kaplan & Kaplan, 1989; Ohly et al., 2016). Alternatively, the increase in frequency of words such as 'beautiful', 'fun', and 'enjoy' during park exposure suggest that individuals may be experiencing an increase in positive emotions and a reduction in stress, as predicted by Stress Reduction Theory (Berto, 2014). While the words 'I' and 'me' do not have an impact on our quantitative analysis due to their neutral sentiment values, there is a distinct decrease in use of these first-person pronouns during park exposure (Fig. 5D). This pattern supports prior work describing nature exposure as an opportunity to shift from an individual to collective mental frame, potentially leading to pro-social behavior (Zhang et al., 2014).

There are also limitations of using Twitter as a platform and we acknowledge our sample of users willing to geolocate may differ from the general population. In 2016, 24% of online adults were active Twitter users, albeit with a slight overrepresentation by younger Americans (Greenwood, Perrin, & Duggan, 2016). Due to the difficulty of extracting this information from Twitter profiles, we were unable to look at how age, gender, and education levels interact with changes in sentiment from park visitation. There is significant variation in how individuals experience and relate to nature; future work should attempt to understand how individual traits mediate the effects of visiting urban greenspace (Gascon et al., 2015). We also recognize that different socioeconomic groups and culturally diverse populations respond differently to conceptions of nature and call for further work teasing apart how varied groups respond to nature contact (Fischer et al., 2018; Frumkin et al., 2017; Maas et al., 2006). Furthermore, cultures in distinct climates will likely demonstrate different relationships with nature exposure – responses to nature contact will likely manifest very differently in a tropical climate compared to San Francisco (Saw, Lim, & Carrasco, 2015).

In this study, we quantified the change in expressed sentiment associated with visits to urban nature by thousands of individuals. In our sample, individuals tweet happier words while visiting parks, and continue to use happier words for several hours following their visit. Tweets posted in Regional Parks, which are larger and greener, are happier than tweets posted in the smaller and less vegetated Civic Plazas and Squares. Based on our word frequency analysis, improved Twitter sentiment from park visits is driven in part by a decline in negative thinking. Our study deepens the evidence base for the mental benefits provided by nature contact in urban areas. As we continue to uncover the psychological mechanisms from nature contact, we can better inform public health policy and target park planning and design to maximize these benefits.

Urban parks can provide restorative environments for people as well as refuge for biodiversity. The benefits of urban nature include many ecosystem services beyond the scope of this study such as storm-water retention and air purification (Elmqvist et al., 2015). Conserving natural spaces and protecting mental health are not typically discussed in the same policy arenas; however, research further linking health with urban greenspace and biodiversity protection can help planners and public health officials build new strategies that support both goals. We suggest building or expanding parks near populations with limited access to greenspace and targeting funds toward the most effective types of parks for mental benefits. With most of the planet's population now living in cities, we must find ways to bring nature to them in a way that supports both biodiversity and human health.

**Data Archiving**
Data from this study is archived on Figshare at: https://figshare.com/projects/2019_Schwartz_TwitterParks/60230